\documentclass[showpacs, twocolumn]{revtex4-1}
\usepackage{graphicx}
\usepackage{amsmath}
\usepackage{appendix}

\begin{document}

\title{Two-dimensional quantum droplets in dipolar Bose gases}

\author{Abdel\^{a}ali Boudjem\^{a}a}
\affiliation{Department of Physics,  Faculty of Exact Sciences and Informatics, Hassiba Benbouali University of Chlef P.O. Box 78, 02000, Ouled Fares, Chlef, Algeria.}

\email {a.boudjemaa@univ-chlef.dz}

\begin{abstract}

We calculate analytically the quantum and thermal fluctuations corrections of a dilute quasi-two-dimensional Bose-condensed dipolar gas.  
We show that these fluctuations may change their character from repulsion to attraction in the density-temperature plane owing to the
striking momentum dependence of the dipole-dipole interactions. 
The dipolar instability is halted by such unconventional beyond mean field corrections leading to the formation of a droplet phase. 
The equilibrium features and coherence properties  exhibited by such droplets  are deeply discussed. 
At finite temperature, we find that the equilibrium density crucially depends on the temperature and on the confinement strength
and thus, a stable droplet can exist only at ultralow temperature due to the strong thermal fluctuations.

\end{abstract}

%\pacs{03.75.Nt, 05.30.Jp} 

\maketitle

\section{Introduction}
%%%%%%%%%%%%%%%%%%%%%%%%%%%%%%%%%%%%%%%%%%%%%%%%%%

One of the most fascinating phenomena recently observed in dipolar Bose-Einstein condensates (BEC) with Dy and Er atoms is the formation of
self-bound droplets \cite {Pfau1, Pfau2, Pfau3, Chom}. 
Theoretically, a wealth of studies have been spawned for highlighting the behavior of  droplet states \cite{ Wach, Bess2, Saito, Wach1, Bess3, BoudjDp, Bess1, Kui, Raf, Mac}.  
These so-called liquid droplets are stable even in the absence of external trapping \cite {Pfau3, Wach1, Bess3} (self-bound) 
due to the competition between attraction, repulsion and Lee-Huang-Yang (LHY) quantum fluctuations \cite {LHY, lime, Boudj1}.
It was also found that dipolar quantum droplets are anisotropic and form a regular array, 
result in from the anisotropy and the long-range character of dipole-dipole interaction (DDI).
These droplets, whose densities are one order of magnitude higher than the density of the ordinary BEC and decay 
at the droplet critical temperature at which particles undergo a phase transition from a low density gas phase  (ordinary BEC) to 
the high density phase (droplet) \cite{BoudjDp}. 

In three-dimensional (3D) case, the LHY corrections provide a term proportional to $n^{3/2}$, where $n$ is the peak density, that arrests the dipolar instability at high condensed density 
\cite{Pfau2, Wach, Bess2, Saito}.
In quasi-1D dipolar BEC, the LHY quantum corrections present anomalous properties due to the transversal modes 
and the quantum droplet is appeared only in a low density regime \cite{Mish}. 
Quantum fluctuations play also a crucial role in stabilizing droplets in low-dimensional nondipolar Bose-Bose mixtures \cite{PetAst, Li, Boudj2}.

In this paper we investigate for the first time the formation of a droplet state in quasi-2D weakly interacting dipolar bosons. 
We show that the system yields many surprising and interesting properties. 
We find that the LHY quantum corrections change their nature from repulsive to attractive 
due to the peculiar momentum dependence  of the DDI \cite {boudjG},  similarly to the quasi-1D droplet \cite{Mish}.
This unconventional behavior not only modifies the density dependence and the quantum stabilization mechanism but also unveils 
novel phase of matter consisting  of a quantum droplet. 
The nucleation of this state occurs due to the competition between the roton instability results in local collapses \cite {boudjG}, and the LHY fluctuations.
It has been suggested that the roton softening combined with the quantum stabilization mechanism opens a new avenue for exploring supersolids \cite{Pfau4}.

%Monte Carlo simulations \cite{Zoller} have revealed that the combination of a repulsive dipolar potential with a short-distance
%cutoff of Rydberg atoms in the dipole-blockade regime may lead to the formation of a supersolid droplet in 2D geometry. 

The observed 2D self-bound dipolar droplets differ from the bound state of 2D weakly attracting bosons \cite {Hum} 
and that of 2D Bose-Bose mixtures with both contact and dipolar interactions  \cite{PetAst, Li, Boudj2}. 
The former exists by the increased kinetic energy associated with their nonuniform shape while the latter 
occurs when the interspecies interaction is weakly attractive and the intraspecies ones are weakly repulsive.
In addition, the transition to the droplet state happens at relatively high density compared to the quasi-1D system \cite{Mish}.
 %Using variational and numerical means we point out that the size of the self-bound decreases exponentially with the number of particles.
The density and the shape of the droplet are analyzed by numerically solving the underlying generalized nonlocal 2D Gross-Pitaevskii equation (GPE). 
We find that the equilibrium density is markedly affected by transversal modes.
It is shown in addition that the LHY corrections invoke non-trivial enhancements in the excitations and the one-body density matrix of the droplet.
Finally, we extend our study to finite temperatures and predict effects of thermal fluctuations.
We determine the stability condition as well as the condensed density inside the droplet. 

The rest of paper is organized as follows.
Sec.\ref{Mod} introduces the DDI in quasi-2D and the LHY quantum corrections.
Sec.\ref{Drop} deals with the stability regime, the ground-state properties and the coherence of the droplet.
In section \ref{Dropth} we generalize our results to finite temperature.
Sec.\ref{Concl} contains our conclusions.

%Conceptually, the droplet structures exhibit peculiar features:
%(i) they are individually Bose condensed and superfluid.
%(ii) They remain self-bound even in the absence of external trapping \cite {Pfau3, Wach1, Bess3}.
%(iii) They decay at temperature close to the transition \cite{BoudjDp}.
%(iv) Their density and shape can be tuned by changing the interaction strength.
%Such liquids and their finite-size droplets remain dilute and weakly interacting allowing for a well-controlled perturbative description. 

%This anomlaous behavior results in
%Moreover, although the condensate remains one-dimensional, the LHY may be crucially affected by 
%transversal modes, which induce a change from attractive to repulsive LHY correction at a critical density.

%%%%%%%%%%%%%%%%%%%%%%%%%%%%%%%%%%%%%%%%%%%%%%%%%%

\section{Model} \label{Mod}

\subsection{ Dipolar interactions in quasi-2D geometry }

We consider a dilute Bose-condensed gas of dipolar bosons tightly confined in the axial direction $z$ by an external potential $U({\bf r})=m\omega^2z^2/2$
and assume that in the $x,y$ plane the translational motion of atoms is free. The dipole moments $d$ are oriented perpendicularly to the $x,y$ plane. 
In the ultracold limit $kr_* \ll 1$, where $r_*=m d^2/\hbar^2$ is a characteristic range of the DDI, 
the momentum representation of the two-body interaction potential $V({\bf r}-{\bf r}')$ is given as \cite{boudjG}
\begin{equation}\label{ddp}
V({\bf k})= g (1- C |{\bf k}|),
\end{equation}
where $C =2\pi d^2/g$, $g=g_{\text{3D}}/\sqrt{2 \pi} l_0$ is the 2D contact interaction coupling constant  
which strongly  depends on the strength of the transverse confinement $l_0=\sqrt{\hbar/m\omega}$, 
and $g_{\text{3D}}=4\pi \hbar^2 a/m$ with $a$ being  the $s$-wave scattering length ($a > 0$ throughout the paper). 
Another model for the effective quasi-2D potential was proposed in Ref.\cite{Nath}
$V(k)= g [1- C kl_0 \exp(k^2l_0^2/2) \text{Erfc} (kl_0/\sqrt{2})]$, where $\text{Erfc}$ is the complementary error function. 
Expanding this potential which can be obtained by integration of the full 3D dipolar interaction 
over the transverse harmonic oscillator at small momenta leads to $V(k)= g (1- C l_0 k)$, 
with $l_0$ adjusts the scale for the strength of the linear term and can be set to unity since the ratio between the dipolar length and the trap length is the most important.
Therefore, both potentials require a high momentum cut-off when calculating the beyond-mean field corrections.
The large momentum behavior of both potentials is different, the potential of  Ref.\cite{Nath} is constant ($-\sqrt{2/\pi}$) for large $k$,
while the potential (\ref{ddp}) is linear in $k$. This implies different regularization schemes when computing the beyond-mean field LHY corrections.

The Bogoliubov excitation energy is given as $\varepsilon_k=\sqrt{E_k^{2}+2 \mu_0 E_k(1-Ck)}$ \cite{boudjG},
where $E_k=\hbar^2k^2/2m$ and  $\mu_0=ng$ is the zeroth order chemical potential. 
For small momenta the excitations are sound waves, $\varepsilon_k=\sqrt{\mu_0/m}k$. 
For $C$ varies as $(\sqrt{8}\xi/3)\leq C\leq \xi$, where $\xi$ is the healing length, the excitation spectrum exhibits a roton-maxon structure \cite{boudjG}.
The observation of such a roton mode has been reported very recently in Ref.\cite{Chom2} 
using momentum-distribution measurements in dipolar quantum gases of highly-magnetic Er atoms.
For $C>\xi$, the uniform Bose gas becomes dynamically unstable.

%\begin{widetext}
%\begin{equation}\label{he2}
%\hat H = \int d^2r \, \hat \psi^\dagger(\mathbf{r}) \left\{\frac{-\hbar^2 }{2m}\Delta_\mathbf{r}+V(\mathbf{r})+
%\frac{1}{2}\int d^2r^\prime\, \hat\psi^\dagger (\mathbf{r^\prime}) V(\mathbf{r}-\mathbf{r^\prime})\hat\psi^\dagger(\mathbf{r^\prime})\right\}\hat\psi(\mathbf{r}) ,
%\end{equation}
%\end{widetext}
%where $\psi^\dagger$ and $\psi$ denote respectevly the usual creation and annihilation field operators, 
%$V(\mathbf{r}-\mathbf{r^\prime})=g\delta(\mathbf{r}-\mathbf{r^\prime})+ d^2 \frac{1-3\cos^2\theta}{\vert r-r^\prime\vert^3}\ $ is the interaction potential
%with $d^2$ is the dipole moment and $\theta$ being the angel between the vector $\vert r-r^\prime\vert$ and the direction of dipoles $(z)$.

%we use the Bogoluibov transformation and separtaing out the condensate from the field operator 
%$\hat\psi=\hat\psi^\prime+\psi_0 $, where $\hat\psi^\prime= \sum_{\vec k} u_k\hat a_k\exp(\frac{-i\varepsilon_k t}{\hbar})-v_k\hat a^\dagger_k \exp(\frac{i\varepsilon_k t}{\hbar})$ 
%accounts for the non-condensed densty.

%%%%%%%%%%%%%%%%%%%%%%%%%%%%%%%%%%%%%%%%%%%%%%%%%%
\subsection{LHY corrections}

Indeed, obtaining reliable estimate for the beyond mean field correction to the equation of state (EoS) in low dimensions is challenging
even for Bose systems with contact interactions \cite{Boudj4, Sal, Zin}.
At zero temperature, the LHY corrections to the EoS can be written \cite{boudjG, BoudjDp}
\begin{equation}    \label{LHY}
\delta \mu_{\text{LHY}}=\frac{1}{2} \int V({\bf k}) \left[\frac{E_k}{\varepsilon_k}-1\right] \frac{d \bf k}{ {(2\pi})^2}.
\end{equation}
The evaluation of this integral requires special care due to the crucial contribution to the beyond mean field terms of the transverse trap modes of the contact interactions.
The large-momentum divergence originating from the dipolar term $-gCk$ (valid only for $k \ll 1/r_*$)  is another issue of the integral (\ref{LHY}).
One possibility to solve this problem is to work with an arbitrary $\Lambda$-cutoff. 
In the case of contact interactions, the potential (\ref{ddp}) takes the form $V(k) = g$ for $k < \Lambda$, and 0 otherwise. 
Then, if $\Lambda$ is larger than typical momenta in the gas, the obtained LHY corrections are cutoff-independent 
and in good agreement with the existing literature (see e.g. \cite{Boudj4, Sal, Zin, GPS}).
Now if one applies this method to the dipolar interaction case, it turns out that the resulting corrections to the EoS
are cutoff-dependent  (the cutoff is not larger than the roton momentum) due to the special character of the DDI (see e.g \cite{Jach}). 
Another possible route to compute the LHY corrections (\ref{LHY}) is to take into account the full transverse structure. 
Obtaining reasonable stable corrections within this technique is also a tedious and time-consuming task 
(diagonalizing the Bogoliubov-De Gennes equations is extremely difficult both analytically and numerically) \cite{Jach}. 

To circumvent this problem, a high-momentum cutoff  is considered here which is valid in the ultracold regime $k \ll 1/r_*$ \cite{boudjG}.
Despite it gives qualitative correct results, it renders much simpler the calculations and captures the main features of the system at hand \cite{boudjG}. 
The choice of this momentum cutoff is not only motivated by computational convenience, but also the obtained corrections 
will be insensitive to the cutoff in contrast to the $\Lambda$-cutoff method.  
After some algebra, we obtain
%Due to these technical difficulties we proceed to a high-momentum cutoff $1/r*$ which is the simpler way to circumvent this problem. 
%The simplest way to circumvent this problem is the use of a high momentum cutoff $1/r_*$ which provides an efficient treatment for the system at hand \cite{boudjG}. 
%The main contribution to $\delta \mu$ comes from $k$ close to $1/r_*$. 
%Using a cut-off regularization method outlined in our recent work \cite{boudjG}, we obtain 
\begin{align} \label {EoS}
\frac{ \delta \mu_{\text{LHY}}}{E_0}& = (4 \pi ^{3/2}/b^2)^2  nr_*^2\bigg \{1-2b (nr_*^2)^{1/2}- 3 b^2 nr_*^2  \\
&+2 b^2 nr_*^2 \ln \left[1/2 (1- b \sqrt{nr_*^2} )  \right]\bigg\}, \nonumber
\end{align}
where $E_0= \hbar^2/m r_*^2$ and $b=  \sqrt{2\pi^{3/2} l_0/ a} $.
In the absence of the DDI, Eq.(\ref{EoS}) excellently agrees with the usual short-range 2D Bose gas EoS (see e.g. \cite{GPS, Boudj4}).
When the roton minimum is close to zero i.e. $C =\xi $, one has 
$\delta \mu_{\text{LHY}}/ E_0 \simeq (8 \pi ^{3/2}/b^2)^2  nr_*^2  \ln\left[1/\sqrt{ b^2 nr_*^2(1-b^2nr_*^2)}\right] $. 
The quantum corrections (\ref{EoS}) are important to halt the collapse of the system when the 
roton touches zero (roton instability).
They can also substantially impact the collective excitations and the thermodynamics of the system.  

\begin{figure}[ htb] 
\includegraphics[scale=0.8]{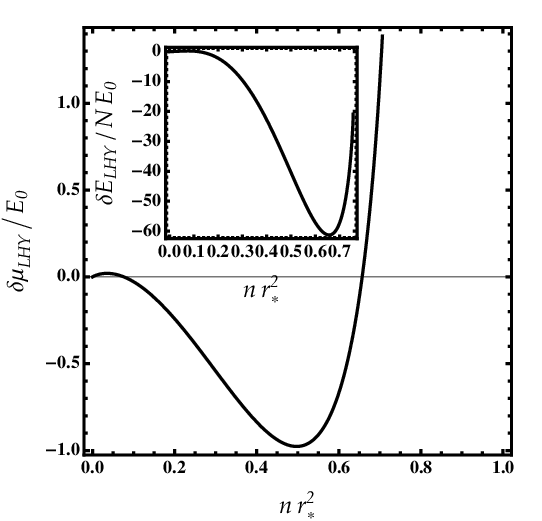}
\caption{ The LHY corrections to the EoS from Eq.(\ref{EoS}) as a function of $nr_*^2$ for  $ l_0/a=40$.
The inset shows the LHY energy.
These parameters are sufficent to reach the roton regime.}
\label{Func}
\end{figure}

Figure \ref{Func} clearly shows that for $0 <n r_*^2 < 0.1$, $\delta \mu_{\text{LHY}}$ initially increases and 
after it reaches its maximum at $n r_*^2=0.05$, starts decreasing. 
In this ultra-dilute limit, $\delta \mu_{\text{LHY}}$ provides an additional repulsive term $\propto n-n^{3/2}$ 
prohibiting the formation of any droplet in contrast to the 2D Bose-Bose mixture with contact interactions \cite{PetAst}.
For $0.1< nr_*^2 \lesssim 0.65$, the effective LHY attraction furnishes an extra term  $\propto - n^2-n^2 \ln (1-n)$ arresting the dipolar instability, 
results in the formation of a stable self-bound droplet.  
This droplet phase has a universal peak density at $nr_*^2 \simeq 0.65$ where the LHY energy, $\delta E_{\text{LHY}}= \int \delta \mu_{\text{LHY}} dn$, 
reaches its minimal value (see the inset of Fig.\ref{Func}).
%This exotic state arises from the competition between the rotonization induced by the DDI and the LHY stabilization.
For  $nr_*^2 > 0.65$, $\delta \mu_{\text{LHY}}$ grows logarithmically and thus, the system undergoes an instability as the complexity increases.

%\begin{align}  \label {FF}
%{\cal F} (C)&=2\left(1-9 \alpha^2+10 \alpha^4\right) \ln \left[ \frac{\beta}{2(1-\alpha)}\right] -\frac{34 \alpha}{3}+ 20 \alpha^3 \nonumber\\ 
%&+ \frac{1}{6 \alpha-3 \beta} \bigg \{\alpha \left[10 \left(\alpha^2-1\right) \beta^2+8 \alpha \beta \left(11-15 \alpha^2\right) \right.  \nonumber \\ 
%&\left. 
%+120 \alpha^2+3\alpha \beta^3-68\right]-6 \beta \bigg \}, 
%\end{align} 

%%%%%%%%%%%%%%%%%%%%%%%%%%%%%%%%%%%%%%%%%%%%%%%%%%
\section{Quantum droplets} \label{Drop}

In this section we discuss the formation and the equilibrium properties of quasi-2D quantum droplets.

The equilibrium density $n_0$ can be obtained by minimizing the energy per particle $\propto E/n$ with respect to the density $n$ \cite{PetAst}, 
where $E=Nng/2+\delta E_{\text{LHY}}$. 
In this manner, we get 
\begin{align}    \label{EqD}
n_0r_*^2 = \frac{\alpha}{b^2}  \exp(1/3b^2),
\end{align}
where $\alpha \simeq 70 \pi/2$. 
Equation (\ref{EqD}) clearly shows that the transverse harmonic confinement $l_0$  may strongly change  the equilibrium density.
Therefore, the weakly interacting regime requires the condition: $b^2 \gg 1$ or equivalently $l_0 \gg a$.

%For instance, the dimensionless equilibrium density $n_0 r_*^2$ is decreasing with increasing $\epsilon_{dd}$. However, as $\epsilon_{dd}$ approaches unity,  
%the mean-field forces dominate the quantum fluctuations at low density and thus, destabilize the droplet state.
%The confinement strength can also change the density $n_0 r_*^2$.

To gain more insights into these quantum ensembles, we numerically solve the generalized  GPE in which $ \delta \mu_{\text{LHY}} \rightarrow  \delta \mu_{\text{LHY}} ({\bf r})$:
\begin{equation}
i\hbar \dot\Phi ({\bf r},t) = \bigg [- \frac{\hbar^2 }{2m} \nabla^2+  {\cal F}^{-1} (V({\bf k})) +\delta \mu_{\text{LHY}} ({\bf r}) \bigg] \Phi ({\bf r},t),  \label{gNLSE}
\end{equation}
where ${\cal F}^{-1}$ is the inverse Fourier transform. 
The wavefunction must satisfy the normalization condition $ 2\pi \int r dr |\Phi |^2=N$. 
In 3D geometry, Eq.(\ref{gNLSE}) has been intensively used to describe the dynamics of the droplet \cite{ Wach, Bess2,  Wach1, Bess3, BoudjDp, Bess1} 
and already validated by quantum Monte Carlo simulations \cite {Saito}.
In the model (\ref{gNLSE}) the effect of higher-momentum modes is just a local density-dependent term and LHY fluctuations
are assumed to be large enough to maintain the overall balance with the dipolar instability. 
The stationary generalized GPE  can be obtained from Eq.(\ref{gNLSE}) using $\Phi ({\bf r},t)=\Phi ({\bf r})\, e^{-i\mu t/\hbar}$, 
where $\mu$ is the chemical potential of the system.  This yields 
\begin{equation}
\mu \Phi ({\bf r}) = \bigg [- \frac{\hbar^2 }{2m} \nabla^2+  {\cal F}^{-1} (V({\bf k})) +\delta \mu_{\text{LHY}} ({\bf r}) \bigg] \Phi ({\bf r}),  \label{sgNLSE}
\end{equation}

\begin{figure}[ htb] 
\includegraphics[scale=0.8] {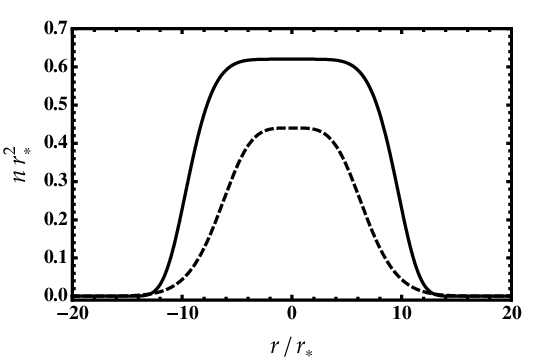}
\caption { Density profiles of  a self-bound droplet.
Parameters are : $ N=10^4$ of ${}^{162}$Dy atoms, $a =141 \,a_0$ \cite{Tang} ($a_0$ being the Bohr radius), and $r_* =130 \,a_0$ \cite {Pfau4,Tanzi}.
Solid line: $b=20$. Dashed line: $b=10$.}
\label{phase}
\end{figure}

The numerical simulation of Eq.(\ref{sgNLSE}) was performed using the split-step Fourier transform and the convolution method to 
evaluate the DDI term \cite{Ron}.  
Our simulations are carried out for ${}^{162}$Dy atoms, with typical atom number $ N=10^4$ 
and $a =141 \,a_0$ \cite{Tang} ($a_0$ being the Bohr radius) which can be controlled via a magnetic Feshbach.
Dy atoms in their ground state have a dipolar length  $r_* =130 \,a_0$ \cite {Pfau4,Tanzi}.  
In this configuration a roton mode is expected to appear.
Figure \ref{phase} shows that a stable droplet is formed due to the competition between the roton instability 
and the LHY  quantum fluctuations and aquires an equilibrium density $n_0 r_*^2 \simeq 0.63$ 
at which the energy develops a local minimum as is foreseen above  (see the inset of Fig.\ref{Func}).
For ${}^{162}$Dy atoms, the stability is reached at densities $n_0 \sim 10^{16}$ m$^{-2}$ and confinement strength $l_0= 2.6 \times 10^{-7}$m.
By further reducing $l_0$, the droplet contracts to a small size (dashed line in Fig.\ref{phase}).

\begin{figure}[ htb] 
\includegraphics[scale=0.43]{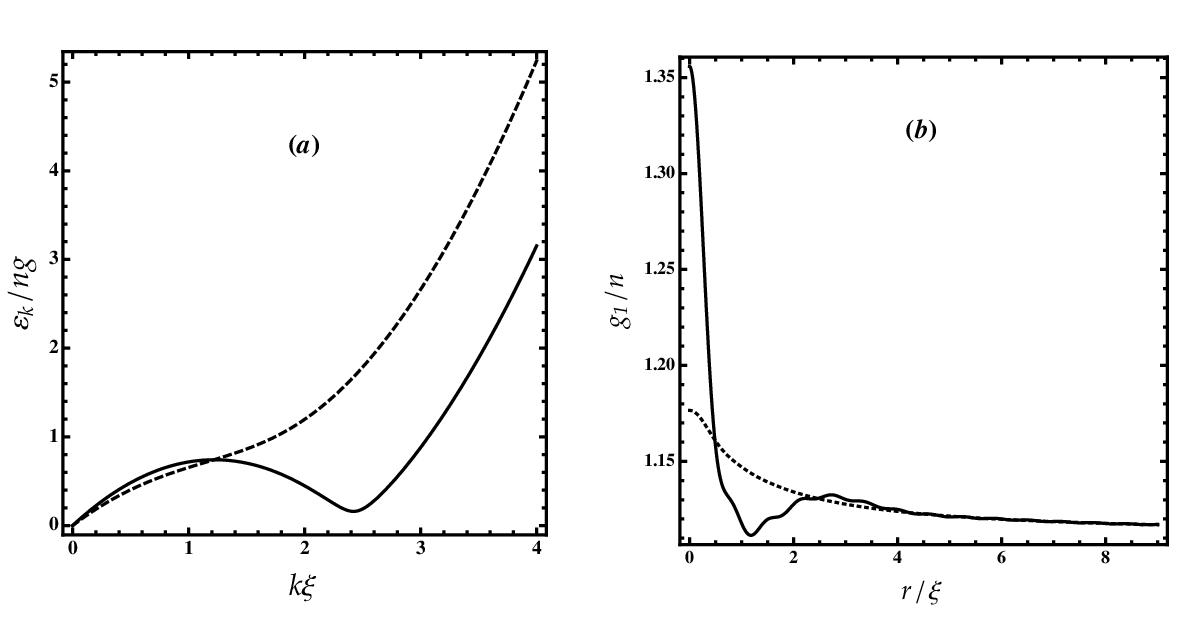} 
\caption{ (a) Excitations energy ${\cal E}_k$ of the quasi-2D dipolar droplet as a function of momentum $k$. Solid line: with LHY corrections and dashed line: without LHY corrections.
 (b) One-body correlation function for $nr_*^2=0.2$ (dotted line) and $nr_*^2=0.75$ (solid line). }
\label{DCF}
\end{figure}

To quantitatively check the existence of the droplet, we additionally analyze the behavior of the one-body density matrix which 
can be determined within the realm of the phase-density representation \cite{pop}. 
Writting the field operator in the form $\hat \psi= \sqrt{ \hat n} e^{i \hat \phi}$, where $\hat \phi$ and $\hat n$ are the phase and density operators, which
obey the commutation relation $[\hat n({\bf r}), \hat \phi ({\bf r'})]=i \delta ({\bf r-  r'})$.
Expanding the density and the phase in the basis of the excitations:
$\delta \hat n ({\bf r}) =\sqrt{n ({\bf r})} \sum_{\bf k} [\sqrt{E_k/{\cal E}_k} ({\bf r})  \hat {b}_{\bf k} +\text{H.C.}]$ 
and $\hat \phi ({\bf r}) =[-i/2\sqrt{n({\bf r}) }]\sum_{\bf k} [ \sqrt{ {\cal E}_k/E_k} ({\bf r}) \hat b_{\bf k}-\text{H.C.}]$ (see e.g \cite{GPS, Boudj4, Boudj5}). 
Assuming small density fluctuations, we then obtain for the excitation spectrum of homogeneous gas
\begin{equation}  \label{specd}
{\cal E}_k=\sqrt{E_k^{2}+2 E_k \mu_0 G(n,b) (1-Ck)},
\end{equation}
where $ G(n,b)=2/b- 3 b \sqrt{nr_*^2}  +4 b^2 n r_*^2 [ \ln 1/(1 - b^2 nr_*^2) -1]$.
Equation (\ref{specd}) shows that the presence of the LHY quantum corrections in the dispersion relation may lead to modify the full spectrum of the system.
If $C$ varies in the narrow interval
\begin{equation}  \label{RM}
\frac{\sqrt{8}} {3}\leq \frac{ C}{ \xi \sqrt{G(n,b)}} \leq 1,
\end{equation}
the system emulates roton-maxon excitation spectrum. 
The position and the gap of the roton are shifted owing to the LHY quantum corrections (see Fig.\ref{DCF}.a) 
in agreement with recent numerical and experimental predictions \cite{Chom2}. 
For $ C>\xi \sqrt{G(n,b)} $, the droplet becomes unstable and thus, the ground state completely disappears.
In the limit $k \rightarrow 0$, the dispersion law (\ref{specd}) is linear in $k$ and well approximated
by the phonon-like linear dispersion form ${\cal E}_k= \hbar c_s k$, where the sound velocity
is given by $c_s=c_0 \sqrt{G(n,b)}$ with $c_0= \sqrt{\mu_0/m}$ being the standard sound velocity. 
The collective excitations of the droplet can be determined by numerically solving the full  Bogoliubov-de-Gennes equations  which are
however, beyond the scope of the present work. 
%the excitation spectrum ${\cal E}_k$ exhibits a roton-maxon structure (see Fig.\ref{DCF}.a). 
%For sufficiently large DDI, the roton gap touches zero and thus, the system undergoes a short-wavelength density modulation.

%It is clear from Fig.\ref{DCF}.b that intriguing behavior is present in the density dependence of the sound velocity at $nr_*^2=0.6$, 
%where $c_s$ has a minimum giving rise to a metastable droplet superfluid phase transition.
%This may open the door to the possibility of observing quantum phase slips \cite{Tanz}.
%For $nr_*^2>0.6$, the sound velocity is sizeable due to the crucial role played by the roton modes.

The one-body density matrix is defined as $g_1({\bf r})= \langle \hat \psi ^{\dagger} ({\bf r}) \hat \psi (0)\rangle=n\exp\{{-\langle [\hat\phi ({\bf r})-\hat\phi (0)]^2]\rangle}/2 \}$.
We see from Fig.\ref{DCF}.b that the one-body density matrix is tending to its asymtotic value $n$ at $r \rightarrow \infty$
signaling the existence of the long-range order allowing formation of a droplet in quasi-2D geometry at zero temperature.
This confirms the scenario anticipated above whereby the combined effect of the quantum fluctuations and the dipolar instability may lead to a stable droplet.
Close to the roton region,  $g_1(r)$ is increased by $\sim 14\%$ and 
exhibits pronounced oscillations when $r$ is approaching to zero.
These oscillations are most likely a signature of the destruction of the long-range order, unlocking the possibility of a novel quantum phase transition.
%This exotic state could be well described within either lattice  or a stripe models \cite{Petrov}.

%At zero temperature, the depletion of the droplet is given by
%\begin{equation}  \label{ncond}
%\tilde n= \frac{1}{2} \int \frac{d \bf k} {(2\pi)^2}   \bigg[\frac{E_k+\mu_0 G(n) (1-Ck)} {{\cal E}_k} -1\bigg].
%\end{equation}
%As is shown in Fig.\ref{DCF}d,  $\tilde n$ initially decreases and then develops a maximum at $nr_*^2 \sim 0.3$.
%For $0.5 < nr_*^2 <0.7$, $\tilde n$ vanishes and hence, the self-bound harbours the total number of particles. % giving rise to a robust self-bound droplet. 
%This result also echo a message from our recent work \cite{Boudj6} that reveals that the LHY depresses the depletion of a disordered droplet.
%For $nr_*^2 >0.7$, the droplet becomes highly depleted. 

%%%%%%%%%%%%%%%%%%%%%%%%%%%%%%%%%%%%%%%%%%%%%%%%%%

\section{Effects of thermal fluctuations} \label{Dropth}

%Now we study a dipolar Bose-condensed gas at finite temperature, in which the particles of the condensate and of the thermal cloud are constrained to move in a plane 
%under radial harmonic confinement and interact via a quasi-2D collisions.  
%Interestingly, a trapped 2D gas, in opposite to the homogeneous gas,  can form a Bose condensate at finite temperature \cite{Hadz, Klep} 
%when the phase coherence length is larger than the Thomas-Fermi radius $l_{\phi}  \gtrsim R_{TF}$ \cite{petr2}. 
%This finite size effects produces by the trapping potential provides a low momentum cut-off for the phase fluctuations and therefore, reduces them.

\begin{figure}[htb] 
\includegraphics[scale=0.65] {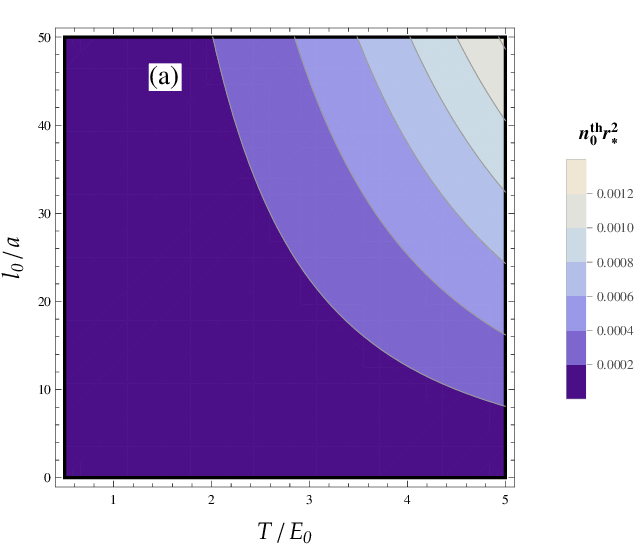}
\includegraphics[scale=0.75] {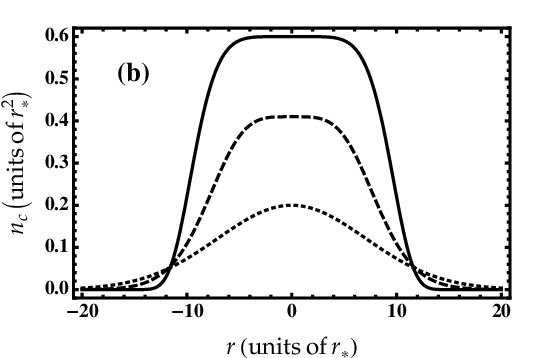} 
\caption { (Color online) (a) Equilibrium density from Eq.(\ref{EqDth}) as function of temperature $T / E_0$ and confinement strength $l_0$.
(b) Condensed density for several values of temperatures $T/E_0$. 
Solid line: $T / E_0=2.5$. Dahsed line:  $T / E_0=3.5$. Dotted line:  $T / E_0=8.5$. 
Parameters are the same as in Fig.\ref{phase}.}
\label{cf}
\end{figure}

In this section we deal with the finite-temperature behavior of the droplet.
In uniform Bose gas, at any nonzero temperature, thermal fluctuations distroy the condensate \cite{merm, hoh}.
However, according to Berezinskii-Kosterlitz-Thouless  (BKT)\cite{Berz, KT},  quasicondensate takes place at low temperature,
characterized by a power-law decay of the one-body spatial correlation function \cite{GPS, Boudj4}. 
In such a quasicondensate, the phase coherence governs only regime of a size smaller than the size of the condensate, marked by the coherence length \cite {GPS}.
For a distance smaller than the coherence length, one can work with the true BEC theory \cite{GPS, Chom1}. 

%To do this, we use the Hartree Fock Bogoliubov method within the Popov approximation. 
%This formalism is a set of non-local equations containing the dipole-dipole interaction and the condensate and thermal correlation functions, which are solved self-consistently.

At finite temperature, the LHY thermal fluctuations reads 
\begin{equation}    \label{deltamuT}
\delta \mu_{\text{LHY}}^{\text{th}}= \int V({\bf k}) \frac{E_k} {\varepsilon_k} [\exp(\varepsilon_k/T)-1]^{-1} \frac{d \bf k} {({2\pi})^2}.
\end{equation}
In contrast to the zero temperature case, integral (\ref{deltamuT}) is finite. At low $T$, the main contribution to Eq.(\ref{deltamuT}) 
comes from the  phonon branch. This yields
\begin{align}    \label{deltamuT1}
\frac{\delta \mu_{\text{LHY}}^{\text{th}}}{E_0}= \frac{b^2} {4 \pi^{3/2}} \bigg[ \frac{\zeta(3)}{\sqrt{2}} (nr_*^2)^{-2} \left(\frac{T}{E_0} \right)^3 \\
- \frac{b }{120 \pi^	2}  (nr_*^2)^{-5/2}\left(\frac{T}{E_0} \right)^4 \bigg], \nonumber
\end{align}
where  $\zeta (3)$ is the Riemann Zeta function. 
%The leading term in Eq.(\ref{deltamuT1}) arises from the short-range interactions while the subleading term accounts for the DDI.  
The most striking feature of the thermal fluctuations (\ref{deltamuT1}) which introduce a new extra term $\propto -n^{-5/2} T^4$,
is that they change their nature from repulsive at lower $T$ to attractive interactions at higher $T$.
Notice that at $T> \mu_0$, the leading term for the chemical potential coincides with that of an ideal gas.

The thermal contribution to the equilibrium density can be given by minimizing 
the free energy $F=Nng/2+T\sum_{\bf k}\ln[1-\exp(-\varepsilon_k/T)]$ \cite{Boudj2, boudjG}.
This yields
\begin{equation}    \label{EqDth}
n_0^{th}r_*^2= \frac{b^2}{7200 \pi^4 \zeta(3)^2}  \left(\frac{T}{E_0} \right)^2,
\end{equation}
its behavior as a function of $l_0$ and $T$ is displayed in Fig.\ref{cf}.a.
We see that the thermal equilibrium density $n_0^{th}r_*^2$ is important only for $T \gg E_0$ and $l_0 \gg a$, revealing that
thermal fluctuations may substantially affact the stability of the droplet. 
At $T \lesssim 2.2 \,E_0$ and $l_0 \lesssim 9 \,a$, $n_0^{th} \ll n_0$ and hence, the condensate is weakly depleted, 
results in the droplet remains in its equilibrium state.

Let us now look at how the condensed density inside the droplet behaves by varying the temperature.
To this end, we insert the quantum (\ref{EoS}) and thermal (\ref{deltamuT1}) fluctuations into the nonlocal GPE (\ref{gNLSE}) and 
pursue typically the same numerical method.
Figure \ref{cf}.b depicts that  the condensed density decreases with increasing  temperature.  
At higher $T$, the droplet evaporates into an expanding gas owing to the strong thermal fluctuations. 
Note that $n_c$ could be also shifted  by changing $l_0/a$ at fixed temperature.

%At finite temperature, the coherence of the droplet is connected to the static structure factor $S({\bf k})$. 
%This later is related to the pair correlation function and defined as $S({\bf k})= \langle \hat n_k \hat n^\dagger_k\rangle/N= (E_k/{\cal E}_k) \coth \left({\cal E}_k/2T\right)$ \cite{Boudj7}.
%As one can see in Fig \ref{cf}.b, as the temperature is increased,  $S({\bf k})$ develops a peak in the region of $k  \simeq 2/\xi$
%indicating that the droplet starts to disappear. 
%One can also easily show that by further increasing $nr_*^2$, the correlations get more and more stronger and the height of the peak is rised.
%This means that in quasi-2D systems, the thermal fluctuations become strong enough to destroy the off-diagonal long-range order 
%and hence preventing the occurrence of the droplet at temperatures $T \gtrsim E_0$.
%At large momenta, $S({\bf k})$ is tending to unity.  

%%%%%%%%%%%%%%%%%%%%%%%%%%%%%%%%%%%%%%%%%%%%%%%%%%

\section{Conclusions} \label{Concl}

In conclusion, we predicted the formation of a self-bound droplet in quasi-2D dipolar Bose gas at both zero and low temperatures. 
Interestingly, the roton instability inducing a local collapse instability can be stabilized by the LHY corrections.
Unlike the Bose mixtures \cite{PetAst},  the LHY quantum and thermal fluctuations which present an intriguing density dependence,
found to be pivotally influenced by the transversal modes. 
Such modes may change the nature of the LHY corrections from attractive to repulsive at certain density. 
Their impacts on the structure of the droplet are also considerable.
The one-body correlation function of the droplet is decaying over distance and displays a remarkable behavior near the roton instability.   
At finite temperature, we pointed out that the droplet state can survive only at ultralow temperatures 
($T < E_0$), that should be smaller than the BKT transition temperature.  
Experimentally, the realization of the quantum droplet remains challenging in particular at finite temperatures due to its self-evaporation.
%The BKT transition and the vortices that accompany could be analyzed by means of Kibble-Zurek mechanism.
One can expect, on the other hand, that the unusual density dependence of the quantum corrections persists also in the presence 
of the three-body correlations \cite{BoudjDp, Dima, Boudj6}.
%The insights obtained in the present study may offer a fresh view of the supersolid droplet.
Future experimental investigations and Monte Carlo  simulation are required 
in order to fully understand the confidentiality of the droplet state in 2D configuration.

%%%%%%%%%%%%%%%%%%%%%%%%%%%%%%%%%%%%%%%%%%%%%%%%%%

\section*{Acknowledgements}
We are grateful to Dmitry Petrov,  Lauriane Chomaz, Krzysztof Jachymski, Grigori Astrakharchik and Pawel Zin
for fruitful discussions and comments on the manuscript. 

\newpage 
\section*{Appendix: Low-energy $S$-wave scattering of dipolar bosons in quasi-2D} \label{A}

In this appendix  we discuss low-energy two-body scattering of identical particles undergoing the 2D translational motion 
and interacting with each other at large separations via the potential 
\begin{equation}    \label{dpot}
V(r)= \frac{d^2}{r^3}=\frac{\hbar^2 r_*}{m r^3},
\end{equation}
where $r_*$ is the characteristic dipole-dipole distance (see the main text). 
The term low-energy means that their momenta satisfy the inequality $kr_*\ll1$. 

The on-shell scattering amplitude is defined as
\begin{equation}    \label{onshell}
f_l (k) =\int_0^{\infty}  J_l(kr) V({\bf r}) \psi_k({\bf r}) \,d{\bf r},
\end{equation}
where $J_l (kr)$ is the Bessel function, and $\psi_k({\bf r})$ is the true wavefunction of the relative motion with momentum $k$. 
It is governed by the Schr\"odinger equation
\begin{equation}  \label{SE}
\left(-\frac{\hbar^2}{m}\Delta+\frac{ d^2} { r^3} \right) \psi_k({\bf r})= \frac{\hbar^2k^2}{m}  \psi_ k({\bf r}), 
\end{equation}
where $m$ is the reduced mass. \\
For the solution of the scattering problem, it is more convenient to normalize the wavefunction of the radial relative motion with orbital angular momentum $l$ 
in such a way that it is real and, for $r \rightarrow \infty$, one has
\begin{equation}  \label{Solu}
\psi_l(r)=  J_l(kr)-\tan \delta_l N_l(kr),
\end{equation}
where $N_l$ is the Neumann function and $\tan \delta_l=-(m/4\hbar^2) f_l (k)$. 
In order to calculate the $s$-wave part of the scattering amplitude, we divide the range of
distances into two parts $r < r_0$ (region I) and $r> r_0$ (region II), where the distance $r_0$ is selected such that $r_* \ll r_0\ll k^{-1}$ \cite{Gora, Boudjthesis}. 
Here we distinguish two contributions to the scattering amplitude :
the short-range contribution coming from distances $r < r_*$, and the so-called anomalous contribution coming from distances of the order of the de Broglie wavelength of particles,
$r \sim k^{-1}$ \cite {Landau}. 

In region I, the $s$-wave relative motion of two particles is governed by the Schr\"odinger equation with zero kinetic energy: 
\begin{equation}  \label{SE1}
\left( \frac{d^2}{dr^2}+\frac{1}{r}\frac{d}{dr}-\frac{ r_*} { r^3} \right) \psi_I (r)=0, 
\end{equation}
which admits the solution 
\begin{equation}  \label{Solu1}
\psi_I (r) \propto \bigg[A K_0 (2\sqrt{r_*/r}+I_0 (2 \sqrt{r_*/r})\bigg],
\end{equation}
where $K_0$ and $I_0$ are the modified Bessel functions and the constant $A$ is determined by the behavior of $V(r)$ at shorter distances.
If $V(r)= d^2/r^3$ at all distances (pure dipole-dipole potential), then $A=0$.
In the case when $V(r)$ behaves as $d^2/r^3$ at distances $r>r_1 \sim r_*$ (square well potential at short distance), then 
the coefficient $A$ can be determined by equalizing the logarithmic derivatives of the wavefunction obtained in the region $r<r_1$ and the one of $\psi_I (r)$
at $r=r_1$. In this model, we should have $A\gg1$, so that $r_d \ll r_*$.

In region II, the relative motion is practically free and the potential $V(r)$ can be considered as perturbation. 
To zero-order we then have for the relative wavefunction
\begin{equation}  \label{Solu1}
\psi_{II}^{(0)} (r)=J_0(kr)-\tan \delta_0^{(0)} N_0(kr),
\end{equation}
where the scattering phase shift $\delta_0^{(0)}$ is due to the interaction between particles in region I.

Matching the logarithmic derivatives of $\psi_I (r)$ and $\psi_{II}^{(0)} (r)$ at $r=r_0$, and taking into account only terms up to $r_*$, we get 
\begin{equation}  \label{Solu3}
\tan  \delta_0^{(0)}= \frac{\pi}{2Q} +\frac{\pi}{2Q^2}  \frac{r_*}{r_0} \chi,
\end{equation}
where
$$Q=\ln \bigg( \frac{k r_d} {2} e^{\gamma} \bigg), \;\;\;\;
\chi= 2+2 \ln \bigg( \frac{r_0}{r_d} \bigg) +\ln^2 \bigg( \frac{r_0}{r_d} \bigg),$$
with $\gamma \simeq 0.557$ being the Euler constant and $r_d=r_* e^{2\gamma-2A}$.

On the other hand, the contributions to the $s$-wave scattering phase shift from distance $r>r_0$ should be included perturbatively. 
In this region, to first-order in $V(r)$, the relative wavefunction is given by
\begin{equation}  \label{Solu4}
\psi_{II}^{(1)} (r) = \psi_{II}^{(0)} (r)- \int_{r_0}^{\infty} G(r,r') V(r') \psi_{II}^{(0)} (r') d{\bf r'},
 \end{equation}
where the Green function for the free $s$-wave motion is given by 
\begin{equation} \label{Solu5}
G(r,r') = -\frac{m}{4\hbar^2}
\begin{cases} 
\psi_{II}^{(0)} (r') N_0 (kr), \;\;\;\;\;\;\;  r >r' & \\
\psi_{II}^{(0)} (r) N_0 (kr'), \;\;\;\;\;\;\;  r<r'
\end{cases}
\end{equation}
Substituting the Green function (\ref{Solu5}) into (\ref{Solu4}) and taking the limit $r \rightarrow \infty$,  we have \cite{Gora, Boudjthesis}
\begin{equation}  \label{Solu6}
\tan \delta_0=\tan \delta_0 ^{(0)}+\tan \delta_0^{(1)}, 
\end{equation}
where the first-order contribution to the phase shift is given by:
\begin{equation}  \label{Solu7}
\tan \delta_0^{(1)}=  2kr_* \left(1+\frac{\pi}{4Q^2} \right)-\frac{\pi}{2Q^2} \frac{r_*}{r_0} \chi.
\end{equation}
The sum $\tan \delta_0^{(0)}$ and the first-order contribution gives
\begin{equation}  \label{Solu8}
\tan \delta_0= \frac{\pi}{2Q} + 2kr_* \left(1+\frac{\pi}{4Q^2} \right).
\end{equation}

Now for identical bosons, the full scattering amplitude  $f_l= -(4\hbar^2/m) \tan \delta_l$,
can be obtained by making summation over all partial amplitudes with even $l$
\begin{equation}  \label{Solu9}
f (k)=-\frac{4\hbar^2}{m} \left[\frac{\pi}{2Q} + 2\pi kr_* \left(1+\frac{\pi}{4Q^2} \right)\right].
\end{equation}
We then omit the term proportional to $kr_*/Q^2$ and notice that in quasi-2D where the inequality $kl_0 \ll 1$ is satisfied,
the parameter $r_d$ depends  on the confinement length $l_0=\sqrt{\hbar/m \omega_0}$ in the $z$-direction 
as $r_d \approx l_0 e^{-l_0/\sqrt{2 \pi} a}$ \cite{GPS,Boudjthesis}.
Substituting this into Eq.(\ref{Solu8}), and keeping in mind that 
in the quasi-2D geometry the short-range constant $g=g_{3D}/(\sqrt{2} l_0)$, we  finally obtain the result (\ref{ddp}) employed in 
the main text, $V({\bf k})= g (1- C |{\bf k}|)$.

%\section*{Author contributions} 

%The author contributed alone to this work.

%\section*{Additional information}
%Competing financial interests: The author declares no competing financial and non-financial interests.


\begin{thebibliography}{28}

\bibitem{Pfau1} H. Kadau, M. Schmitt, M. Wenzel, C. Wink, T. Maier, I. Ferrier-Barbut and T. Pfau, Nature {\bf 530} ,194 (2016).
\bibitem{Pfau2} I. Ferrier-Barbut, H. Kadau, M.Schmitt, M. Wenzel, T. Pfau, Phys. Rev. Lett. {\bf 116}, 215301, (2016).
\bibitem {Pfau3} M.Schmitt, M. Wenzel, F.B\"ottcher, I. Ferrier-Barbut and T. Pfau, Nature {\bf 539}, 259 (2016).
\bibitem{Chom} L. Chomaz, S. Baier, D. Petter, M. J. Mark, F. W\"achtler, L. Santos and F. Ferlaino, Phys. Rev. X {\bf 6}, 041039 (2016). 

\bibitem{Wach} F. W\"achtler and L. Santos, Phys. Rev. A {\bf 93}, 061603 (R) (2016).
\bibitem {Saito} H. Saito, J. Phys. Soc. Jpn. {\bf 85}, 053001 (2016).
\bibitem{Bess2} R. N. Bisset R. M. Wilson D. Baillie and P. B. Blakie, Phys. Rev. A {\bf 94}, 033619 (2016).
\bibitem{Wach1} F. W\"achtler and L. Santos, Phys. Rev. A {\bf 94}, 043618 (2016).
\bibitem{Bess3} D. Baillie, R. M. Wilson, R. N. Bisset, and P. B. Blakie, Phys. Rev. A  {\bf 94}, 021602(R) (2016).
\bibitem{BoudjDp}  A. Boudjem\^{a}a, Annals of Physics, {\bf 381}, 68 (2017).
\bibitem{Bess1} R. N. Bisset and P. B. Blakie, Phys. Rev. A {\bf 92}, 061603(R) (2015).
\bibitem{Kui} Kui-Tian Xi and Hiroki Saito, Phys. Rev. A {\bf 93}, 011604(R) (2016).
\bibitem{Raf}  R. Ołdziejewski and K. Jachymski, Phys. Rev. A {\bf 94}, 063638 (2016).
\bibitem{Mac} A. Macia, J. S\'anchez-Baena, J. Boronat, and F. Mazzanti, Phys. Rev. Lett. {\bf 117}, 205301 (2016).
\bibitem{LHY} T. D. Lee, K. Huang and C. N. Yang, Phys. Rev {\bf 106}, 1135 (1957).
\bibitem {lime} Aristeu R. P. Lima and Axel Pelster, Phys. Rev. A {\bf 84}, 041604 (R) (2011); Phys. Rev. A {\bf 86}, 063609 (2012).
\bibitem{Boudj1} A. Boudjem\^{a}a, J. Phys. B: At. Mol. Opt. Phys.  {\bf 48}, 035302 (2015); J. Phys. A: Math. Theor. {\bf 49}, 285005 (2016).
\bibitem {Mish} D. Edler, C. Mishra,  F. W\"achtler, R. Nath, S. Sinha, and L. Santos,  Phys. Rev. Lett. {\bf 119}, 050403 (2017).
\bibitem{PetAst} D. S. Petrov and G. E. Astrakharchik,  Phys. Rev. Lett. {\bf 117}, 100401 (2016).
\bibitem{Li} Y. Li, Z. Luo, Y. Liu, Z. Chen, C. Huang, S. Fu, H. Tan and B. Malomed, New J. Phys. {\bf 19}, 113043 (2017). 
\bibitem{Boudj2} A. Boudjem\^{a}a, Phys. Rev. A {\bf 98}, 033612 (2018). 
\bibitem{Nath} R. Nath, P. Pedri, and L. Santos, Phys. Rev. Lett. {\bf 102}, 050401 (2009).

\bibitem {boudjG} A. Boudjem\^{a}a and G. V. Shlyapnikov, Phys. Rev. A {\bf 87}, 025601 (2013); Abdel\^{a}ali Boudjem\^{a}a, Phys.Lett.A, {\bf 379} 2484 (2015).
\bibitem{Pfau4}  F. B\"ottcher, J-N. Schmidt, M. Wenzel, J. Hertkorn, M. Guo, T. Langen, and T. Pfau, Phys. Rev. X {\bf 9}, 011051 (2019).
%\bibitem{Zoller} F. Cinti, P. Jain, M. Boninsegni, A. Micheli, P. Zoller, and G. Pupillo, Phys. Rev. Lett. {\bf 105}, 135301 (2010).
\bibitem{Hum} H.-W. Hammer and D. T. Son, Phys. Rev. Lett. {\bf 93}, 250408 (2004).
\bibitem{Chom2}  L. Chomaz, R.M. W. van Bijnen, D. Petter, G. Faraoni, S. Baier, J. Hendrik Becher, M. J. Mark, F. W\"achtler, L. Santos, F. Ferlaino, Nat. Phys. {\bf 14}, 442 (2018);
                            D. Petter, G. Natale, R. M. W. vanBijnen, A. Patscheider, M. J. Mark, L. Chomaz and F. Ferlaino, Phys. Rev. Lett. {\bf 122}, 183401 (2019).
\bibitem{Boudj4}  A. Boudjem\^{a}a, Phys.Rev.A. {\bf 86}, 043608 (2012).
\bibitem{Sal} L. Salasnich and F. Toigo, Phys. Rep. {\bf 640}, 1 (2016).
\bibitem{Zin} P. Zi\'n, M. Pylak, T. Wasak, M. Gajda, Z. Idziaszek,  Phys. Rev. A {\bf 98}, 051603 (2018).
\bibitem{GPS}  See for review: D.S. Petrov, D.M. Gangardt, and G.V. Shlyapnikov, J. Phys. IV (France) {\bf 116}, 5 (2004).
\bibitem{Jach}  K. Jachymski and R. Ołdziejewski, Phys. Rev. A {\bf 98}, 043601 (2018).
\bibitem{Ron}  S. Ronen, D. C. E. Bortolotti, and J. L. Bohn, Phys. Rev. A {\bf 74}, 013623 (2006).
\bibitem{Tang} Y. Tang, W. Kao, K.-Y. Li, S. Seo, K. Mallayya, M. Rigol, S. Gopalakrishnan, and B. Lev, Phys. Rev. X {\bf 8}, 21030 (2018).
\bibitem{Tanzi} L. Tanzi, E. Lucioni, F. Fam\'a, J. Catani, A. Fioretti, C.Gabbanini, and G. Modugno, Phys. Rev. Lett. {\bf 122}, 130405 (2019).
\bibitem{pop} V.N. Popov, {\it Functional Integrals in Quantum Field Theory and Statistical Physics} (D. Reidel Pub., Dordrecht, 1983).
\bibitem{Boudj5}  A. Boudjem\^{a}a, Phys.Rev. A {\bf 94}, 053629 (2016).
\bibitem{merm} N. D. Mermin, and H. Wagner, Phys. Rev. Lett. {\bf 22}, 1133 (1966).
\bibitem{hoh} P. C. Hohenberg, Phys. Rev. {\bf 158}, 383 (1967).
\bibitem{Berz} V. L. Berezinskii,  Soviet Phys. JETP {\bf 34}, 610 (1971).
\bibitem{KT} J.M. Kosterlitz and D.J. Thouless, J.Phys. C {\bf 6}, 1181 (1973); J.M. Kosterlitz, J. Phys. C {\bf 7}, 1046 (1974).
\bibitem{Chom1} L. Chomaz, L. Corman, T. Bienaim\'e, R. Desbuquois, C. Weitenberg, S.  Nascimbène, J. Beugnon, J. Dalibard, Nature Communications {\bf 6}, 6172 (2015).
\bibitem{Boudj7}  A. Boudjem\^{a}a,  J. Phys. B: At. Mol. Opt. Phys. {\bf 49}, 105301 (2016).
\bibitem{Dima}  D. S. Petrov, Phys. Rev. Lett. {\bf 112}, 103201  (2014).
\bibitem{Boudj6}  A. Boudjem\^{a}a, J. Phys. B: At. Mol. Opt. Phys. {\bf 51}, 025203 (2017).

\bibitem{Gora} J. Levinsen, N. R. Cooper and G. V. Shlyapnikov, Phys. Rev. A {\bf 84}, 013603 (2011).
\bibitem{Boudjthesis} A. Boudjem\^{a}a, Dynamics of ultracold gases, Ph.D. thesis, Hassiba Benbouali University of Chlef, (2013).
\bibitem{Landau} L. D. Landau and E. M. Lifshitz, Quantum Mechanics (Butterworth-Heinemann, Oxford, 1999).

\end{thebibliography}
\end{document}